\documentstyle[12pt]{article}
\font\blackboard=msbm10 % scaled \magstep1
\font\blackboards=msbm7 \font\blackboardss=msbm5
\newfam\black \textfont\black=\blackboard
\scriptfont\black=\blackboards \scriptscriptfont\black=\blackboardss

\font\ninerm=cmr9
\def\uniset{\rlap{\ninerm 1}\kern.15em 1}

\newcommand{\be}{\begin{equation}}
\newcommand{\ee}{\end{equation}}
\newcommand{\bea}{\begin{eqnarray}}
\newcommand{\eea}{\end{eqnarray}}

\title{{\bf Distribution of Husimi Zeroes in Polygonal Billiards}}
\author{{\bf Debabrata Biswas} \\
{\it Theoretical Physics Division }\\
{\it Bhabha Atomic Research Centre} \\
{\it Trombay, Mumbai 400 085, India} \\
{E-mail : dbiswas@apsara.barc.ernet.in} \\
\\
{\bf Sudeshna Sinha} \\
{\it The Institute of Mathematical Sciences} \\
{\it  CIT Campus, Taramani } \\
{\it Chennai 600 113, India} \\
{E-mail : sudeshna@imsc.ernet.in}}

\begin{document}
\maketitle {\abstract The zeroes of the Husimi function provide a
minimal description of individual quantum eigenstates and their
distribution is of considerable interest. We provide here a numerical
study for pseudo-integrable billiards which suggests that the zeroes
tend to diffuse over phase space in a manner reminiscent of chaotic
systems but nevertheless contain a subtle signature of
pseudo-integrability. We also find that the zeroes depend sensitively
on the position and momentum uncertainties ($\Delta q$ and $\Delta p$
respectively) with the classical correspondence best when $\Delta q =
\Delta p = \sqrt{\hbar/2}$. Finally, short range correlations seem
to be well described by the Ginibre ensemble of complex matrices.

\section{Introduction}
\label{sec:Intro}

This paper deals with phase space parameterizations of one-dimensional
{\em billiard map} eigenfunctions for polygonal
enclosures. Specifically, we shall deal with the Bargman-Husimi
representation and study the distribution of its zeroes for regular,
irregular and bouncing ball modes. Such a study has been carried out
before for integrable and chaotic billiards \cite{tualle_voros,leb_vor_95}
and these systems are now reasonably well understood in the sense that the
distribution reflects a correspondence with the underlying classical
dynamics.  As with most other objects of interest in generic polygonal
(pseudo-integrable) billiards, the distribution of zeroes is
interesting if only to explore the existence of such a correspondence
with the classical system.

Of all possible Hamiltonian systems, billiards are perhaps the best
understood category and exhibit the entire gamut of classical dynamics
depending on the shape of the enclosure.  Of these, polygonal
billiards form an important sub-category and apart from the rectangle
and the triangles $(\pi/3,\pi/3,\pi/3)$, $(\pi/2,\pi/3,\pi/6)$,
$(\pi/2,\pi/4,\pi/4)$, all other polygonal enclosures are
non-integrable \cite{pjr_mvb}.  Further, the ones with rational
interior angles are pseudo-integrable; they have two constants of
motion as in integrable systems and yet their invariant surface in
phase space has a genus, $g > 1$. One of the simplest examples of a
pseudo-integrable system is the $\pi/3$ enclosure for which $g = 2$
{\it i.e. the invariant surface is a double torus}. Here, as in other
pseudo-integrable billiards, an initial (parallel) beam of
trajectories splits after successive encounters with the $2\pi/3$ (in
general $m\pi/n, m > 1$) vertex and traverse different paths.

There are several important consequences of pseudo-integrability at
the classical level that are now known. However, as far as
semiclassics is concerned, pseudo-integrable billiards are still
rather poorly understood. When the dynamics is integrable, an EBK
ansatz for the wavefunction \cite{keller}
\be
\psi (q) \sim \sum_{j=1}^N A_j\exp(i S_j/\hbar)
\label{eq:EBK}
\ee

\noindent
works well at least in the limit $\hbar \rightarrow 0$.  In the above,
$S_j$ are the (finitely many) branches of the classical action at
energy $E$ and $A_j$ are constant amplitudes for integrable
polygons. Such an ansatz however does not work for pseudo-integrable
billiards even though the number of sheets that constitute the
invariant surface is still finite. We shall not discuss the reasons
for its breakdown here but merely remark that no definite behaviour
for pseudo-integrable eigenfunctions is known. For classically chaotic
systems on the other hand, the Schnirelman theorem \cite{schnirel}
(suitable phase-space measures constructed from the eigenfunctions
must tend towards the classical phase-space ergodic measure as $\hbar
\to 0$) does provide a semiclassical constraint albeit in a measure
theoretic sense. Besides, there exist results on the amplitude
distribution and spatial correlation function which have been subject
to tests \cite{mcdonald88}.

Despite the absence of any such result for pseudo-integrable polygons,
numerical studies \cite{db90} such as those for the amplitude
distribution or nodal plots suggest that typical eigenfunctions are
irregular and broadly speaking, there is little to distinguish them
from the eigenfunctions in chaotic systems. In the present paper, we
shall try to refine this existing body of knowledge and will employ
for this purpose a phase-space representation of quantum mechanics,
which is known to highlight certain semiclassical features for
integrable and chaotic systems.  Our results are empirical, based on
extensive numerical studies and can be simply expressed as follows :
the eigenfunctions of polygonal billiard as viewed in the Husimi
representation tend to be irregular {\em but nevertheless contain a
subtle signature of classical pseudo-integrability}.

The paper is organized along the following lines.
In section~\ref{sec:Formalism}, we briefly review the Husimi - Bargman
representations and the results on random analytic functions.
We introduce the systems that we shall study and the quantum
map under consideration in section~\ref{sec:models}.
This is followed by our numerical results on the Husmini function and
the density of zeroes in section~\ref{sec:numerics}. Finally,
correlations are discussed in section~\ref{sec:correlations}
and our conclusions are summarized in section~\ref{sec:Conclusions}.

\section{Phase Space Representations}
\label{sec:Formalism}

Phase space representations of quantum wavefunctions are best suited
in semiclassical studies since the quantum dynamics (Heisenberg
equation) then appears as an explicit deformation of the classical
dynamics (Liouville equation) by shifting the analysis onto the
density operator $\rho = | \psi > < \psi |$. In quantum mechanics
however, the phase space representation of a state is not unique
since operators ($\hat{q}, \hat{p} $ for instance ) may
be ordered in various ways while having the same classical analog.
A general expression for a quasi-probability distribution function may
be expressed as \cite{ZFG}

\be
\rho_{(\Omega )}(q,p,t) = {1\over (2\pi)^2}\int~d^2\xi~e^{i(\xi^*z^*
+ \xi z) \hbar }~{\rm Tr}~[ \Omega\{e^{-i\xi^*\hat{a}\dagger}
e^{-i\xi\hat{a}} \} \hat{\rho} ] \label{eq:basic}
\ee

\noindent
where $\Omega$ refers to the ordering that is chosen. The Wigner
distribution follows from a symmetric ordering of ($\hat{q}, \hat{p} $)
which implies

\be
 \Omega\{e^{-i\xi^*\hat{a}\dagger}
e^{-i\xi\hat{a}} \} = e^{-i\xi^*\hat{a}\dagger - i\xi\hat{a}}
\label{eq:weyl}
\ee

\noindent
while the Husimi function is a result of anti-normal ordering

\be
 \Omega\{e^{-i\xi^*\hat{a}\dagger}
e^{-i\xi\hat{a}} \} = e^{-i\xi\hat{a}}e^{-i\xi^*\hat{a}\dagger}
\label{eq:antinormal} .
\ee

\noindent
Using Eqns.~(\ref{eq:basic}) and (\ref{eq:weyl}),
the distribution function in the Wigner representation,
$\rho_w(q,p;\hbar)$, for a pure state can be explicitly written
as :

\be
\rho_w(q,p;\hbar)  = {1\over (2\pi\hbar)^d} \int
< q - \eta / 2 | \psi > < \psi | q + \eta / 2 >
e^{ip.\eta/\hbar} d\eta
\label{eq:wigner}
\ee

\noindent
where $d$ is the degree of freedom of a dynamical system. Thus, the
expectation of a dynamical variable $\hat{A}$ is represented as

\be
{\rm Tr}~[~\hat{A}~| \psi > < \psi |~] = \int A_w(q,p) \rho_w(q,p)~dqdp
\label{eq:trace}
\ee

\noindent
where

\be
A_w(q,p) = \int < q - \eta / 2 |~\hat{A}~| q + \eta / 2 >
e^{ip.\eta/\hbar} d\eta
\ee

The Wigner function however takes positive as well as {\em negative}
values and oscillates violently with a wavelength $\hbar$ in phase
space. A coarse grained distribution function is thus preferred
and the Husimi function,

\be
\rho_h(q,p;\hbar) = {1\over (\pi\hbar)^d} \int \rho_w(q',p';\hbar)
e^{-\sum_{i=1}^{N}[{(q_i-q'_i)^2 \over 2(\Delta q_i)^2} +
{(p_i-p'_i)^2] \over 2(\Delta p_i)^2}]} dp'dq'
\label{eq:husimi}
\ee

\noindent
is one such example which can be expressed as a smoothened Wigner
function. In this case, the smoothening is achieved through the
Gaussian centred at a phase space
point $(q,p)$. In Eq.~(\ref{eq:husimi}) above,

\be
\Delta q_i = \sqrt{ {\hbar \over 2} } \sigma_i,~~~~~\Delta p_i = \sqrt{ {\hbar \over 2} } {1 \over \sigma_i}
\label{eq:uncertain}
\ee

\noindent
are the uncertainties in $q$ and $p$ respectively.  Note that $\rho_h$
is merely a minimum-uncertainty (m.u.) state decomposition of the
wavefunction $\psi$ and can be expressed as

\be
\rho_h(q,p;\hbar) = { | < z | \psi > |^2 \over 2\pi\hbar }
\label{eq:husimi_coh}
\ee

\noindent
where

\be
| z >  = e^{ {-|z|^2\over 2}} \sum_{n=0}^\infty
{z^n\over \sqrt{n!}} | n >
\ee

\noindent
$ \{ | n > \} $ are the harmonic oscillator number states,
$a^\dagger = (\sigma^{-1/2} \hat{q} - \imath \sigma^{1/2} \hat{p})/
(\sqrt{2\hbar})$ and $ z = (\sigma^{-1/2} q - \imath \sigma^{1/2} p)/
(\sqrt{2\hbar})$ with $\sigma > 0$.
Note that $ < z | z > = 1 $ while $ < z | z' > \neq 0 $.
Written explicitly for $1~-$ degree of freedom,

\be
<x|z> = \left ( {1\over 2\pi (\Delta q)^2} \right )^{1/4} e^{ipx~-~
{(x - q)^2 \over 4 (\Delta q)^2}}
\ee

\noindent
which is the minimum uncertainty wavepacket whose Wigner transform is
the Gaussian used in Eq.~(\ref{eq:husimi}).

From Eq.~(\ref{eq:husimi_coh}), it is evident that
$\rho_h$ takes only positive values. The minimum wavepackets, $ | z >
$ and $ < z | $ are eigenfunctions of $\hat{a}$ and $\hat{a}^\dagger$
respectively with eigenvalues $z$ and $z^*$.
Eq.~(\ref{eq:husimi_coh}) follows directly from eqns.~(\ref{eq:basic})
and (\ref{eq:antinormal})
using the expansion of the identity operator

\be
\hat{{\rm I}} = \int d\mu(z)~|z><z|
\ee

\noindent
where $d\mu(z) = dqdp/(2\pi\hbar)$.

If the system under consideration is ergodic, the Husimi density $
\{\rho_h^{n}\} $, corresponding to a sequence of eigenstates $
\{\psi_n(q)\} $ with eigenvalues $ E_n \rightarrow E$, almost always
converges to the classical Liouville measure $\mu_E$ over the energy
surface $\Sigma_E$.  Thus, if $f(q,p)$ is any smooth observable,

\be
\int~f(q,p)~\rho_h^{n}~dqdp \rightarrow \int_{\Sigma_E}~f(q,p)~
d\mu_E~~~{\rm as}~~~E \rightarrow E_n.
\label{eq:ergodic}
\ee

\noindent
Schnirelman's theorem however allows an occasional exception
(e.g. scarred state) and for this reason, a more appropriate
description of non-integrable eigenfunctions is desirable.

In 1990, Leboeuf and Voros \cite{le_vo_jphys_a_90} proposed that
the zeroes of the Husimi function provide a minimal
description of quantum states \cite{more_recent}. The first step in this
direction is the coherent state ($\sigma = 1$) or Bargman representation,
$ < z | \psi > $ of a state $ | \psi > $ which
maps unitarily the standard Hilbert space onto the space of
{\em entire} functions with finite Bargman norm

\be
\parallel \psi \parallel = {1 \over 2\pi\hbar} \int_{R^2} | \psi(z) |^2
e^{-|z|^2 } dq dp.
\label{eq:barg_norm}
\ee

\noindent
One can thus consider $\psi(z)$ as a phase phase representation of the
wavevector $ | \psi >$.  Note that the zeroes of the Bargman and
Husimi functions are identical. The Bargman function however contains
information about the phase (of the wavefunction) as well and is hence
a more fundamental object.
For the standard case when the phase space is a plane
(the Weyl - Heisenberg group, $W_1$),

\be
\psi(z)  =  e^{ {-|z|^2\over 2}} \sum_{n=0}^\infty {a_n\over
\sqrt{n!}}~z^n ~~~~~~~~~~~~~~~~~~~~~(W_1)  \label{eq:W_1}
\ee

\noindent
where $a_n$ are the expansion coefficients of $ | \psi > $ in
terms of the harmonic oscillator number states. Similar results
can be written down for the sphere (~$SU(2)$~) and the pseudo-sphere
(~$SU(1,1)$~) \cite{perelomov,leboeuf_recent} though unlike the
case of $W_1$ or $SU(1,1)$, the Bargman representation
of $| \psi >$ for $SU(2)$ is finite reflecting the compactness
of phase space. For Hamiltonian systems however, energy conservation
does ensure that the manifold is compact so that Eq.~(\ref{eq:W_1})
has, in practice, only a finite number of terms. Clearly then, the
Husimi-Bargman zeroes specify a state completely.

It is evident that the distribution of the Husimi-Bargman zeroes
depends on the distribution of the expansion coefficients $\vec{a}
= (a_1,a_2,\ldots,a_n)$.
For chaotic systems, it is natural to expect that the choice of an
arbitrary basis (harmonic oscillator in this case) makes $\vec{a}$
point in any direction of Hilbert space with equal probability
\cite{leboeuf_recent}. The only constraint then comes from
normalization so that $\sum a_n^2 = 1$. For purposes of computing
the distribution of zeroes, this is equivalent to the
assumption that the coefficients are drawn from a Gaussian
distribution \cite{kac}

\be
D(\vec{a}) = {1\over (2\pi)^N}~e^
{- \sum_i {|a_i|^2 \over 2} }
\label{eq:gauss}
\ee

\noindent
Eq.~(\ref{eq:W_1}) with the above distribution
is referred to as a {\em random analytic function}.

Random analytic functions (RAF) for various groups  have
been studied in some detail when the coefficients are
complex \cite{leboeuf_recent,bogo_bohi_lebo,
leb_shukla,hannay} corresponding to systems without time
reversal symmetry. The results point to a universal behaviour.
Thus, the density of zeroes is uniform with spacings of
the order of $1/\sqrt N$  and the 2-point
correlation has a simple form \cite{hannay,hannay_all}
independent of the location
of the zeores. Importantly, random analytic functions do seem
to model chaotic systems very well \cite{leb_shukla,shukla}.

For RAF with real coefficients (systems
with time reversal symmetry), Prosen \cite{prosen} has
studied the density and the k-point correlations. The density
in this case is non-uniform due to the presence of zeroes
on the symmetry axis (the real line). Away from the real
axis however, the density becomes uniform and in this region,
correlations tend towards the case with complex coefficients.
There are few numerical studies however on chaotic systems
with time reversal symmetry though it might be expected
that RAF with real coefficients do model them rather
well.

In contrast, it is known \cite{voros_pra} that for integrable systems,
eigenfunctions follow a WKB-type Ansatz (see eq.~\ref{eq:EBK} ) in the
Bargmann representation too, from which it follows that the zeroes lie
on fixed curves which are anti-Stokes lines of the complex classical
action in the $z$ variable, along which the zeros are equi-spaced with
the separation of order $1/N$.

For the sake of completeness, it may also be noted that
a random polynomial

\be
\psi(z) = a_0 + a_1z + a_2Z^2 + \ldots + a_Nz^N
\label{eq:ranpoly}
\ee

\noindent
with coefficients distributed according to Eq.~(\ref{eq:gauss}),
has zeroes which tend to accumulate around the unit circle
\cite{bogo_bohi_lebo}.

With this background, we shall explore the distribution of Husimi
zeroes for polygonal billiard eigenfunctions in the following
sections. Unless otherwise stated, we shall consider enclosures with
unit perimeter and $\sigma = 1$ (coherent state). We shall also
consider the energy, $E = 1$ and instead quantize $\hbar$ so that
$\hbar = 1/k$. The $\hbar \rightarrow 0$ then corresponds to the
classical dynamics at $E = 1$.

\section{Polygonal Billiards and the Quantum Map}
\label{sec:models}

Classical billiards are enclosures within which a point particle
undergoes specular reflection. The dynamics thus depends on its
shape. For rational polygonal enclosures, the dynamics is constrained
by two constants of motion such that the invariant surface is
two-dimensional.  For the rectangle and the integrable triangles, this
is a torus for which $g=1$. For all other rational polygons, the
invariant surface is topologically equivalent to a sphere with
multiple holes ($g > 1$).  The simplest example is a double torus (g =
2) which corresponds to enclosures such as the $\pi/3$ rhombus or the
L-shaped billiard. In general, the genus of any rational polygon can
be calculated from its interior angles. Thus, if $m_i\pi/n_i$ are the
interior angles of a rational polygon,

\be
g = 1 + {N\over 2} \sum_i {m_i - 1\over n_i}
\label{eq:genus}
\ee

\noindent
where $N$ is the least common multiple of $n_i$ so that the number of
sheets that constitute the invariant surface is $2N$. Thus various
sets of internal angles may have the same genus but with different $N$
such that the number of distinct momenta spanned by a generic
trajectory varies from enclosure to enclosure.

While the genus does affect certain classical features of the
system \cite{db_pramana},
its influence on quantum states is not known for certain. Studies on
irrational and rational rhombus billiards show that there is little
difference between the morphologies of generic eigenfunctions or their
Husimi densities \cite{shudo_shimizu_95}.  Shudo and Shimizu
\cite{shudo_shimizu_95} even note
that ``~\ldots the difference between random features of
eigenfunctions of quantum polygonal and the desymmetrized dispersing
system are minute \ldots ''. The only difference, they noted, was the
occurrence of bouncing ball states though these can be observed in
other chaotic systems such as the Stadium billiard.

Our investigation of polygonal billiard eigenfunctions lies in this
backdrop. Instead of the Husimi densities themselves, we shall study
their zeroes following Tualle and Voros \cite{tualle_voros}.
The systems we choose
are triangles and rhombus billiards and for all practical purposes,
these can be treated as pseudo-integrable systems irrespective of the
internal angle \cite{see_hobson,high_genus}.

The eigenvalues and eigenfunctions can be obtained by solving the
Helmholtz equation

\be
(\nabla^2 + E) \Psi(q) = 0
\label{eq:helm}
\ee

\noindent
with $\Psi(q) = 0$ on the boundary. The problem can however be reduced
to an eigenvalue problem for an integral operator $K$ or a {\em
Quantum Poincare Map} in various ways \cite{boasman} :

\bea
\psi(s) &  = & \oint ds' \psi(s') K_D(s,s';k) \\
K_D(s,s';k) & = & - {\imath k \over 2} \cos \theta(s,s') H_1^{(1)}
(k|\vec{s} - \vec{s'}|) \\
\cos \theta(s,s') & = & \hat{n}(\vec{s}).\hat{\rho}(s,s')
\label{eq:bim}
\eea

\noindent
where $E = k^2$, $\hat{\rho}(s,s') =
(\vec{s} - \vec{s'})/|\vec{s} - \vec{s'}|$
and $ \hat{n}(\vec{s})$ is the outward normal at the point $\vec{s}$.
The unknown function is now the normal derivative
on the boundary
\be
\psi(s) = \hat{n}(\vec{s}).\nabla \Psi(\vec{s})
\label{eq:psi_Psi}
\ee

\noindent
and the full interior eigenfunction can be recovered through the
mapping

\be
\Psi(q) = - {\imath \over 4} \oint ds H_0^{(1)}(k|\vec{s} - \vec{s'}|) \psi(s)
\label{eq:Psi_psi}
\ee

\noindent
Thus, the essential dynamical information lies within the reduced
$1-d$ function $\psi(q)$ and we shall use this to study phase space
representations and look at their zeroes.
For an enclosure of unit perimeter (which we
shall assume from now on) $ \psi(q + 1) = \psi(q)$. The Bargman
transform, $\psi(z)$ thus obeys a quasi-periodicity condition as well
\cite{tualle_voros} :

\be
\psi(z + 1) = e^{{i \over \hbar}p}~\psi(z)
\ee

\noindent
and the norm-finiteness condition becomes :

\be
\parallel \psi \parallel = {1 \over 2\pi\hbar} \int_{-\infty}^{+\infty}
~dp \int_0^1~dq~|\psi(z)|^2 e^{-|z|^2}~~<~~\infty
\ee

\section{Husimi Zeroes in Polygons - Results}
\label{sec:numerics}

The distribution of Husimi-Bargman zeroes in polygonal
billiards has not been investigated before and as remarked earlier,
the only properties known about the eigenfunctions are from numerical
studies. The lack of concrete results leaves us with little
expectation and perhaps the only conjecture that can be made is that
the distribution of Husimi zeroes of polygonal billiards should differ
from the regularly spaced zeroes along fixed curves typical of
integrable systems.

Note that classical Poincare section plots in suitable (Birkhoff)
co-ordinates do not immediately reveal the dramatic difference between
integrable and pseudo-integrable polygons. In both cases, the points
lie along a finite number of $\sin \theta = {\rm constant}$ lines
where $\theta$ is the angle between the ray and the inward normal at
the boundary point $q$.  Thus there is little difference between the
Poincare sections of the equilateral triangle and the $\pi/3$
rhombus. With increasing genus however, the number of such lines
generally increase as the trajectory explores larger number of
momentum directions.

Semiclassically, the Husimi eigen-distribution function is known to be
localized near the torus for integrable systems \cite{takahashi} while
its zeroes distribute themselves along curves maximally distant from
the invariant curves (anti-Stokes lines). As an example, we first
consider the equilateral triangle billiard.  Fig.~1 shows the Husimi
distribution of a typical eigenstate with quantum number $(m,n) =
(26,81)$ while Fig.~2 is a plot of its zeroes. Clearly, the Husimi
distribution is peaked on the corresponding torus as evident from
Fig.~3 while the zeroes lie on lines located away from the
torus. Further, the zeroes are equispaced on each line though the
spacings typically do vary from line to line.

The zeroes do not always distribute themselves along
straight lines in all integrable polygons and the equilateral triangle
with its high symmetry is a rather special case. In fact, the
distribution of zeroes of an
equilateral state viewed in another enclosure (~related by symmetry -
for instance the ($\pi/6,\pi/3,\pi/2$) triangle or the $\pi/3$
rhombus~) looks very different. Fig.~4 is an example where the
fixed curves are not always straight lines though the zeroes are
equi-spaced along each curve.

As examples of pseudo-integrable polygons, we shall consider rhombus
and triangle billiards. Since the choice of enclosure plays an
important role in determining the distribution of zeroes, we shall use
the $\pi/3$ rhombus to compare the regular and irregular states. Note
that the regular states in this case correspond to equilateral
triangle modes which vanish on the shorter diagonal and they comprise
approximately half the total number of states in the $\pi/3$
enclosure (fig.~4 is an exmaple).
The irregular states on the other hand are ``pure rhombus''
modes \cite{db90} which do not vanish on the shorter diagonal. Barring
the bouncing-ball modes, ``pure rhombus'' modes display features
typical of irregular wavefunctions. We shall look for the differences
in the distribution of zeroes between (i) regular and irregular modes
and (ii) bouncing-ball and non-bouncing-ball ``pure rhombus'' modes.

Fig.~5 displays the zeroes of a
typical irregular ``pure rhombus'' mode. The zeroes are no longer
distributed along curves and they tend to diffuse all over the phase
space. Note that there is a reflection symmetry in this case about the
$q=0.25, 0.5$ and $0.75$ lines so the zeroes need only be viewed in a
quarter of the phase space. Clearly, they are more or less randomly
distributed with no clear alignment along any curve barring some
exceptions where two or more zeroes are distributed around some $p =
{\rm constant}$ line. These observations are in sharp contrast to the
distribution of zeroes for integrable polygons.

We next look at the zeroes of a neighbouring bouncing ball
state. Studies on the stadium billiard have shown that the Husimi
zeroes of bouncing ball modes are distributed randomly over the entire
phase space as in case of irregular modes - an observation that may
seem counter-intuitive keeping in mind the existence of approximate
quantum numbers in the description of such states
\cite{bai_hose_etal}.  Fig.~6 shows the Husimi zeroes of a typical
bouncing ball mode in the $\pi/3$ rhombus. The distribution is no
different from the earlier case with few zeroes distributed around $p
= {\rm constant}$ lines and the other zeroes distributed randomly.

The symmetry of the rhombus leads to redundant zeroes and hence poor
statistics as compared to an unsymmetric polygon at the same
energy. However, it does show that the Husimi zeroes do not align
themselves along fixed curves but rather tend to diffuse over phase
space with some amount of clustering around a few $p =$ constant
lines.  As further evidence, we display the Husimi zeroes of a typical
state in the ($\pi/4,\pi/5$) triangle in Fig.~7. They are indeed
distributed over the entire classical phase space while the dashed
lines indicate a tendency to cluster around certain momenta.  This
effect however seems to be pronounced only in systems with low
genus. Thus for the triangle with internal angles
($97\pi/301,79\pi/501$), there seems to be little or no
clustering (see Fig.~8) and the zeroes seem to be genuinely
distributed over the entire phase space as in chaotic billiards.
Fig.~9 shows a set of four histograms which illustrate this
difference in clustering. The $x$ axis of the histograms give the
momenta value and the $y$ axis shows the fraction of zeros occurring in
a bin. The peaked distribution at specific $p$
values for the low genus ($\pi/4,\pi/5$) triangle indicates a
clustering of its zeros. In contrast the high genus case shows an
almost uniform distribution of zeros away from the real axis
marked by a nearly flat histogram (barring the enhanced
density around $p=0$).

Thus, eigenstates of generic \cite{high_genus}
pseudo-integrable billiards tend to behave like
their chaotic counterparts insofar as the distribution of zeroes is
concerned. This suggests that there is no obvious semiclassical
correspondence in non-integrable polygonal billiards. In integrable
polygons however, the correspondence is clear at least when $ \Delta
q_i = \Delta p_i = \sqrt{\hbar/2} $ (see Eq.~(\ref{eq:uncertain}).
However, when this not so (the minimum uncertainty state is not a
coherent state \cite{coherent}), the zeroes tend to move with
$\sigma$.  As an example, we display here the zeroes of an equilateral
triangle mode for two values of $\sigma$ in Fig.~10. When $\sigma =
\sqrt{\hbar/2}$, the zeroes are equi-spaced and lie on a
line. However, as $\sigma$ is reduced, the zeroes move outwards and
realign themselves on a curve as shown in the figure. Finally, as
$\sigma$ is reduced further, the zeroes start moving out of the
classical phase space

\section{Correlations}
\label{sec:correlations}

In the previous section, we found that the zeroes in non-integrable
polygonal enclosures are uniformly distributed away from the real
axis and hence are like those of random analytic functions  with
real coefficients which presumably model chaotic systems with time
reversal symmetry. To ascertain how close the distributions are,
we shall study here the nearest neighbour spacings distribution, $P(s)$
and the 2-point correlation, $R_2(r)$.

\subsection{Nearest Neighbour Distribution}

The nearest neighbour spacings distribution is the simplest
statistic to perform though there exists no analytic predictions
for RAF with real or complex coefficients. The curve in Fig.~11
for random analytic function is thus determined numerically.
A total of approximately 25,000 zeroes from 50 eigenstates
of three different non-integrable triangles has been used
for computing the nearest neighbour distribution of generic
polygons. The zeroes have been unfolded such that
$\int~s~P(s)~ds = 1$

Fig.~11 shows a plot of the integrated spacings distribution,
$I(s) = \int_0^s~P(s') ds'$ for polygons and a comparison with
random analytic function having real coefficients. The agreement
is fair but  there are deviations indicating perhaps that the
underlying assumption about the distribution of coefficients
(see Eq.~(\ref{eq:gauss}) is not fully justified.

Remarkably however, the Ginibre ensemble \cite{mehta,haake}
of complex random matrices shows much better agreement as evident from
fig.~11. In this case, the integrated spacings distribution \cite{haake},
$I_G(s) = i(<s> s)$ where $<s> =
\int_0^\infty ds [ 1 - i(s) ] = 1.142929$ and

\be
i(s) = 1 - \lim_{N \rightarrow \infty} \prod_{n=1}^{N-1}~[ e_n(s^2)~e^{-s^2} ]
\ee

\noindent
where

\be
e_n(x) = 1 + {x \over 1!} + {x^2 \over 2!} + \ldots + {x^n \over n!}
\ee

\noindent
At small values of $s$, $I_G(s) \sim s^4$ and hence $P(s) \sim s^3$.
In comparison, the nearest neighbour spacing distribution
for uncorrelated points thrown at random on the plane
exhibits no level repulsion.

\subsection{Two-point correlation}

For $SU(2)$ random analytic functions with complex coefficients,
the k-point correlation  function has been computed by
Hannay analytically. In particular, the 2-point function,
$R_2({\bf r_1},{\bf r_2}) = <~\rho({\bf r_1})~\rho({\bf r_2})~>$
depends only on the relative distance $r$ between points
${\bf r_1}$ and ${\bf r_2}$ since the density if uniform.
In the asymptotic (number of zeroes, $N \rightarrow \infty$)
limit

\be
R_2(r) \simeq { (~\sinh^2~v + v^2~)~\cosh~v  - 2v~\sinh~v \over
\sinh^3~v } \label{eq:R_2_complex}
\ee

\noindent
where $v = \pi r^2/2$ and $r$ is measured in terms of the
mean spacing ( $\sqrt{4\pi/N}$ for the sphere ). This result holds for other
phase space topologies as well when $N \rightarrow \infty$
and the coefficients are complex.

For systems with time reversal symmetry (real coefficients),
the density is not uniform everywhere and hence
$R_2({\bf r_1},{\bf r_2})$
is sensitive to the location of the zeroes. Away from the real
axis however, $R_2$ has the limiting behaviour given by
Eq.~(\ref{eq:R_2_complex}).

For the Ginibre ensemble of complex random matrices, the density
is uniform and the two
point correlation (in unfolded units)

\be
R_2({\bf r_1},{\bf r_2}) = 1 - \exp(- \pi\left| r_1 - r_2 \right|^2)
\label{eq:R_2_ginibre}
\ee

\noindent
is a function of the distance between the two zeroes.
Note that Eq.~(\ref{eq:R_2_ginibre}) does not have the
characteristic hump at $r \simeq 1$ associated with random
analytic functions.

In Fig.~12, we present results for three different triangles. The
close agreement
suggests that there is possibly a universality in the
distribution of zeroes of non-integrable polygons (corroborated
by similar studies on the nearest neighbour).
We next compare the average of the combined data with the
predictions for the Ginibre ensemble (~see Eq.~(\ref{eq:R_2_ginibre})~)
and Eq.~(\ref{eq:R_2_complex}). The deviations from the RAF
predictions~\cite{fnote_r2}
are evident while the Ginibre ensemble result agrees with our
data very well.

\section{Conclusions}
\label{sec:Conclusions}

We have studied the distribution of Husimi zeroes in polygonal
billiards in this paper and our observations can be summarized as
follows :

\vskip 0.1 in
$\bullet$
In integrable enclosures, the Husimi density is peaked on the
classical torus and the zeroes lie equi-spaced on fixed curves that
are located away from the torus when the minimum uncertainty state is
a coherent state.

\vskip 0.1 in
$\bullet$
The zeroes tend to move as the uncertainties in position and momentum
are varied even as they obey the minimum uncertainty relation. Thus,
coherent states ($\Delta p = \Delta q = \sqrt{\hbar/2} $ ) are the
most classical of all minimum uncertainty states.

\vskip 0.1 in $\bullet$ A weak signature of pseudo-integrability can
be associated with the clustering of some zeroes around a few lines as
observed in some low-genus polygons.

\vskip 0.1 in
$\bullet$
For generic pseudointegrable enclosures, the zeroes tend to be randomly
distributed over the entire phase space as in chaotic billiards
or random analytic functions with real coefficients. This
is especially true for polygons with high genus.

\vskip 0.1 in
$\bullet$
The nearest neighbour spacings distribution of zeroes and the
two point correlation, $R_2(r)$, suggests
that for pseudo-integrable billiards,
the correlations are very well described by the
Ginibre ensemble of complex random matrices. It
is however not clear why this is so and a proper understanding
is desirable.

\section{Acknowledgements}

The authors acknowledge stimulating discussions with
Prof. A.~Voros and thank Dr. Pragya Shukla for valuable help in
our studies on correlations. D.B also acknowledges several
useful discussions on quasi probabilty distributions with
Dr. R.~R.~Puri.

\pagebreak
\centerline{\bf Figure Captions}
\bigskip

\noindent
Fig. 1. The $z$-axis is the Husimi density, $\rho_h$,
of the equilateral triangle
eigenfunction at $ k = 900.142$ corresponding to the
quantum numbers $(26,81)$. \\

\noindent
Fig. 2. Husimi zeroes of the eigenfunction considered in
Fig.~1. \\

\noindent
Fig. 3. Classical surface of section plot in Birkhoff co-ordinates
of trajectories on the corresponding torus (see Fig.~1 for details).
Here $ p = \sin \theta $ where $\theta$ is the angle between the ray
and the inward normal at the point $q$ measured along the
boundary. \\

\noindent
Fig. 4. Husimi zeroes of the equilateral triangle mode at
$k =  532.751$ viewed in the $\pi/3$ rhombus. \\

\noindent
Fig. 5. Zeroes of an irregular ``pure rhombus'' mode at
$ k = 650.336$. \\

\noindent
Fig. 6. Zeroes of a neighbouring bouncing-ball ``pure rhombus'' mode at
$ k = 660.531$. \\

\noindent
Fig. 7. Husimi zeroes of the $(\pi/4,\pi/5)$ triangle mode at
$900.239$.  The dashed lines indicate constant $p$ lines about which
some zeroes have a tendency to cluster. \\

\noindent
Fig. 8. Husimi zeroes of the ($97\pi/301,79\pi/501$) triangle
for which $g >> 1$. The distribution is similar to those of chaotic
billiards even though the invariant surface is $2$-dimensional.
Here k = 1500.1803. \\

\noindent
Fig. 9. Histograms illustrating the clustering of zeros. The $x$ axis
gives the value of momenta $p$ ($p \in [0,1]$ is shown; the interval
$p \in [-1,0]$ is a reflection of $[0,1]$). The $y$ axis of the
histogram shows the fraction of zeros occurring in a bin around
$p$. Here bin size is $0.025$. Four cases are shown: (a) zeroes of the
$(\pi/4,\pi/5)$ triangle mode at $600.099$; (b) zeroes of the
$(\pi/4,\pi/5)$ triangle mode at $900.239$; (c) zeroes of the
($12345\pi/89762,2011\pi/5431$) triangle at $600.125$ and (d) zeroes
of the ($12345\pi/89762,2011\pi/5431$) triangle at $900.071$. The
peaked distribution at specific (almost equidistant) $p$ values for
the low genus ($\pi/4,\pi/5$) triangle indicates a clustering of its
zeros. In constrast the high genus case shows an almost uniform
distribution of zeros marked by a nearly flat histogram (barring the
enhanced density around $p=0$).\\

\noindent
Fig. 10. Two sets of zeroes for $\Delta q = \sqrt{\hbar/2} = 0.0645832
\left ( \times \right )$ and $0.018 (\diamond) $ respectively for an
equilateral state.  Notice the zeroes moving away in the latter
case. As $\Delta q$ is reduced further, some of the zeroes leave the
classical phase space. \\

\noindent
Fig.~11. The integrated nearest neighbour distribution, $I(s) =
\int_0^s P(s') ds' $ for (i) non-integrable triangles (short-dashed curve)
(ii) Gaussian random analytic function with real coefficients (long-dashed)
and (iii) the prediction for the Ginibre ensemble of complex matrices
(solid line). \\

\noindent
Fig.~12. The two-point correlation, $R_2$ for three different
non-integrable triangles ($8\pi/31,17\pi/97$), ($97\pi/301,79\pi/501$)
and ($12345\pi/89762,2011\pi/5431$). Here $r$ is measured in terms
of the mean spacing $\sqrt{2\pi\hbar}$. \\

\noindent
Fig.~13. The two-point correlation, $R_2$ averaged over the three
triangles (histogram) compared to the prediction for Gaussian random
analytic functions with complex coefficients (broken curve) and
the Ginibre Ensemble prediction (~solid curve - see
Eq.~(\ref{eq:R_2_ginibre})~).


\begin{thebibliography}{99}
\bibitem{more_recent}~More recently, Nonnenmacher and
Voros \cite{non_vor} have rigorously
established that the semiclassical convergence of the Husimi densities
towards the uniform ergodic measure implies the equidistribution of
the corresponding zeroes over the phase space in the classical
limit.
\bibitem{hannay_all} In \cite{hannay}, Hannay has derived a
general expression for k-point correlation function.
\bibitem{high_genus} For computational purposes, irrational polygons
are systems with high genus.
\bibitem{fnote_r2} Note that in the Figs.~11 and 13 ,
no attempt has been made
to separate the zeroes close to and far away from the real
axis for either the RAF result on $I(s)$ or the polygonal
billiard results on $I(s)$ and $R_2(r)$. The effect of non-uniformity
has however been found to be small for the number of zeroes considered.
\end{thebibliography}
\end{document}